\documentclass[final,3p,times,twocolumn]{elsarticle}
\usepackage{lineno,hyperref}

\usepackage{amsmath}
\usepackage{graphicx,epsfig,wrapfig}
\usepackage[caption=false]{subfig}

\usepackage{amsthm,mathrsfs}
\usepackage{makeidx}
\usepackage{amsfonts}
\usepackage{amssymb}
\usepackage{amsmath}
\usepackage{mathtools}
\usepackage{fixmath}
\usepackage{amsbsy}
\usepackage[capitalize]{cleveref}
\usepackage{mathrsfs}
\usepackage{slashed}
\usepackage{bm}
\usepackage{multirow}
\usepackage{booktabs}
\usepackage{adjustbox}

\usepackage{color}
\usepackage[normalem]{ulem}

\definecolor{myGreen}{rgb}{0.2,0.72,0.2}
\renewcommand\sout{\bgroup \color[rgb]{0.55,0.00,0.99} \ULdepth=-.5ex \ULset}

\usepackage{simplewick}

\usepackage{multirow}
\usepackage{braket}

\newcommand{\ta}{\left(}
\newcommand{\qa}{\left[}

\newcommand{\tc}{\right)}
\newcommand{\qc}{\right]}

\newcommand{\vet}[1]{\bm{{#1}}}

\renewcommand{\[}{\begin{equation}}
\renewcommand{\]}{\end{equation}}
\allowdisplaybreaks

\modulolinenumbers[5]
\journal{Physics Letters B}
\biboptions{sort&compress}
\begin{document}
\begin{frontmatter}

\title{The gravitational form factor $D(t)$ of the electron}
\author[andreasmainaddress]{Andreas Metz}
\address[andreasmainaddress]{Department of Physics, SERC, Temple University, Philadelphia, PA 19122, USA}  
\author[mymainaddress,mysecondaryaddress]{Barbara Pasquini}
\author[mymainaddress,mysecondaryaddress]{Simone Rodini\corref{mycorrespondingauthor}\fnref{fn1}}
\cortext[mycorrespondingauthor]{Corresponding author}
\ead{simone.rodini@unipv.it}
\fntext[fn1]{\textit{Phone number:} +39 0382 98 7447}
\address[mymainaddress]{Dipartimento di Fisica, Universit\`a degli Studi di Pavia, I-27100 Pavia, Italy}
    
\address[mysecondaryaddress]{Istituto Nazionale di Fisica Nucleare, Sezione di Pavia, I-27100 Pavia, Italy}


\begin{abstract}
The electron-graviton interaction can be described in terms of the gravitational form factors of the QED energy momentum tensor. 
Here we focus on the form factor $D(t)$, and we examine its properties and its interpretation in terms of internal forces at one-loop accuracy in QED. 
We perform the calculation by keeping separate the contributions due to the electron and the photon parts of the energy-momentum tensor.
We also study the case of a nonzero photon mass.
Furthermore, we discuss similarities with and differences to the form factor $D(t)$ of hadronic bound states.
\end{abstract}

\date{\today}


\end{frontmatter}

\section{Introduction}
\label{sec_introduction}

The matrix elements of the energy-momentum tensor (EMT) embody fundamental information about a system~\cite{Kobzarev:1962wt, Pagels:1966zza}. 
They can be parametrized in terms of gravitational form factors (GFFs) that allow one to access the distributions of energy, momentum, orbital angular momentum and internal forces. 
Among them, the GFF $D(t)$, with $t$ indicating the squared momentum transfer to the target, is very intriguing as it appears in the parametrization of the matrix elements of the stress tensor and as such defines the ``mechanical properties" of a system. 
This becomes particularly appealing when applied to hadrons~\cite{Polyakov:2002yz,Polyakov:2018zvc,Lorce:2018egm,Freese:2021czn,Lorce:2015lna}, since it opens a new avenue to unravel 
their underlying quark and gluon structure
as explored in various models~\cite{Ji:1997gm,Polyakov:2018exb,Lorce:2017wkb,Schweitzer:2019kkd,Schweitzer:2002nm,Goeke:2007fq,Goeke:2007fp,Cebulla:2007ei,Jung:2014jja,Kubis:1999db,Belitsky:2002jp,Pasquini:2007xz,Pasquini:2014vua,Mai:2012yc,Mai:2012cx,Cantara:2015sna,Granados:2019zjw,Neubelt:2019sou,Hudson:2017xug,Alharazin:2020yjv,Azizi:2019ytx,Azizi:2020cfc,Ozdem:2020ieh}, lattice QCD~\cite{Hagler:2009ni,Shanahan:2018nnv,Shanahan:2018pib}, and experimental analysis~\cite{Burkert:2018bqq,Kumericki:2019ddg,Dutrieux:2021nlz,Burkert:2021ith}.

The GFFs are also of fundamental importance for the electron, as is the case for the electromagnetic form factors.
We recall that radiative corrections in quantum electrodynamics (QED) generate a nonzero Pauli form factor $F_2(t)$ for the electron, where $F_2(0) = \alpha/2\pi$ is Schwinger's famous one-loop result for the anomalous magnetic moment of the electron~\cite{Schwinger:1948iu}.
Likewise, a nonzero $D(t)$ for the electron is generated by QED loop corrections, with the one-loop calculation first reported in Ref.~\cite{Berends:1975ah} --- see also Ref.~\cite{Milton:1976jr}. 
However, the long-range nature of QED leads to the interesting result that $D \equiv D(t = 0)$, the so-called D-term~\cite{Polyakov:1999gs}, is divergent, while $D(t \neq 0)$ is finite~\cite{Berends:1975ah, Milton:1976jr}.
The form factor $D(t)$, and its Fourier transform in position space, have also been studied using effective field theory by focusing on the region of small $|t|$ (or large distances)~\cite{Donoghue:2001qc}.
The potential impact of the QED long-distance contribution on $D(t)$ of the proton has been highlighted 
recently~\cite{Varma:2020crx}.
Furthermore, the QED GFFs of the electron have attracted new interest in order to explore its angular momentum~\cite{Brodsky:2000ii, Ji:2015sio} and mass structure~\cite{Rodini:2020pis}.
In this work, we extend the previous one-loop QED calculations of $D(t)$ by considering the separate contributions arising from the electron and the photon parts of the EMT.
We also explore the case of a nonzero photon mass which, in particular, leads to a finite D-term.
Moreover, we transform the results to position space which provides the distribution of pressure and shear forces in the electron.
We also point out similarities and differences between the one-loop QED results for $D(t)$ of the electron and (strongly interacting) bound states.
Finally, we discuss the so-called mechanical radius of the electron.

The paper is organized as follows: In Sec.~\ref{sec_definitions} we introduce the EMT in QED, including the extension to a nonzero photon mass, along with the GFFs which parametrize the EMT matrix elements between electron states. 
In this section, we also briefly describe the one-loop QED Feynman diagrams entering the calculation of the GFFs.
We present the (numerical) results for $D(t)$ in Sec.~\ref{FormFactorsEMTSection}, while in Sect~\ref{PressureSection} we examine the distributions of the pressure and shear forces.
In Sec.~\ref{SummarySection} we summarize our main findings.

\section{Definitions}
\label{sec_definitions}
We begin by recalling the (symmetric) Belinfante-Rosenfeld EMT in QED~\cite{Belinfante1,Belinfante2,Rosenfeld}:
\begin{align}
T_{\rm QED}^{\mu\nu} &= T_{e}^{\mu\nu} + T_\gamma^{\, \prime \mu\nu} \,, \;\, \textrm{with} 
\label{T_QED} \\
T_{e}^{\mu\nu} &= Z_2 \,\bar\psi \, \frac{i}{4} \, \gamma^{\{\mu} \overset{\leftrightarrow}{\partial}\phantom{\partial}\hspace{-0.25cm}^{\nu \}}
\psi - \frac{Z_2}{2} \, \mu^{2 \, \varepsilon_{\text{\tiny{UV}}}} \, e \, \bar \psi \, \gamma^{\{\mu} A^{\nu\}} \psi \,, 
\label{Te} \\
T_\gamma^{\, \prime \mu\nu} &= -Z_3 \, F^{\mu\alpha}F^\nu_{\ \alpha} + Z_3 \, \frac{g^{\mu\nu}}{4}F^{\alpha\beta}F_{\alpha\beta} \,, 
\label{Tgamma_prime}
\end{align}
where  the labels $e$ and $\gamma$ refer to the electron and photon contributions, respectively, and $a^{\{\mu}b^{\nu\}} \equiv a^\mu b^\nu+a^\nu b^\mu$.
In Eqs.~(\ref{Te}) and~(\ref{Tgamma_prime}), all the fields and the elementary charge $e$ are renormalized, with 
contributions proportional to $(Z_{2,3} - 1)$ 
representing standard Lagrangian counterterms. 
To deal with the ultraviolet divergences we have used dimensional regularization in $d= 4 - 2 \, \varepsilon_{\text{\tiny{UV}}}$ dimensions with the mass scale $\mu$.

For the calculation with a nonzero photon mass $m_\gamma$ we must go beyond $T_{\rm QED}^{\mu\nu}$ in Eq.~(\ref{T_QED}).
There exist several extensions of QED in order to incorporate a massive photon.
The two most important ones are the Stueckelberg Lagrangian and the spontaneous breaking of the U(1) gauge symmetry, that is, the so-called Abelian Higgs model.
(For a review of both of them see Ref.~\cite{Ruegg:2003ps}.) 
In general, the two approaches describe different theories, but they are equivalent for the purpose of an $O(\alpha)$ calculation of the EMT electron matrix elements.
In fact, both extensions lead to a EMT of the form
\begin{equation}
T^{\mu\nu}_{\rm QED} + m_\gamma^2 \, \bigg( A^\mu A^\nu - \frac{g^{\mu\nu}}{2}A^2 \bigg) + T^{\mu\nu}_{\rm extra} \,,
\end{equation}
and provide identical results since the contribution from $T^{\mu\nu}_{\rm extra}$ (which differs in the two cases) vanishes at $O(\alpha)$.
The same holds for any QED extension with a nonzero $m_\gamma$.

We have worked with the Abelian Higgs model which is defined through the Lagrangian
\begin{displaymath}
\mathcal{L}_{\rm Higgs}^{\rm U(1)} = \mathcal{L}_{\rm QED} + D_\mu \Phi^{\dagger} D^\mu \Phi - g \bigg( \left|\Phi\right|^2 -\frac{v^2}{2} \bigg)^2 \,.
\label{Higgs_Lagrangian}
\end{displaymath}
In Eq.~(\ref{Higgs_Lagrangian}), $D_{\mu}$ is the covariant derivative and $\Phi$ the complex scalar (Higgs) field with the nonzero vacuum expectation value 
\begin{equation}
\braket{0 | \, \Phi \, | 0} = \frac{v}{\sqrt{2}} \,.
\end{equation}
The EMT for the spontaneously broken theory can be written as 
\begin{equation}
T^{\mu\nu}_{\rm QED} + e^2v^2 \, \bigg( A^\mu A^\nu - \frac{g^{\mu\nu}}{2}A^2 \bigg) + T^{\mu\nu}_{\rm Higgs} \,,
\end{equation}
where $T^{\mu\nu}_{\rm Higgs}$ is the EMT for the Higgs sector, which contains also the interaction terms between the Higgs field and the gauge field. 
In the following, we interpret $ev$ as the photon mass.
Since in this simple model the Higgs field is not coupled directly to the matter part of the QED Lagrangian, all the contributions from $T^{\mu\nu}_{\rm Higgs}$ to the matrix elements of the EMT between electron states are $O(\alpha^2)$ or higher. 
Therefore, $T^{\mu\nu}_{\rm Higgs}$ does not contribute in our calculation.
For $v \to 0$ one obtains the massless-photon limit.
More precisely, in this limit one recovers the standard QED for an electron, plus a scalar charged massless particle with a quartic self-interaction --- see Eq.~\eqref{Higgs_Lagrangian} for $v = 0$.

The EMT of interest for our calculation is therefore given by 
\begin{align}
T^{\mu\nu} &= T_{e}^{\mu\nu} + T_\gamma^{\mu\nu} \,, \;\, \textrm{with}
\\
T_\gamma^{\mu\nu} &= T_\gamma^{\, \prime \mu\nu} + m_\gamma^2 \, \bigg( A^\mu A^\nu - \frac{g^{\mu\nu}}{2}A^2 \bigg) \,,
\label{Tgamma}
\end{align}
where we have ignored the Lagrangian counterterms.
All the calculations are performed in the so-called $R_1$-gauge, in which the gauge-fixing term of the Lagrangian takes the form
\begin{displaymath}
\mathcal{L}_{\rm g.f.} = -\frac{1}{2} \ta \partial\cdot A - m_\gamma\phi_2\tc^2 \,,
\end{displaymath}
where $\phi_2$ is the imaginary component of the Higgs field. 
In this gauge, the photon propagator reads (see, e.g., Ch.~3 of Ref.~\cite{Itzykson:1980rh} and Ref.~\cite{Munster:2003gq})
\begin{equation}
iD^{\mu\nu}(k) = \frac{-ig^{\mu\nu}}{k^2-m_\gamma^2+i\epsilon} \,.
\end{equation}

\begin{figure*}[t]
   \centering
   {\includegraphics[width=.15\textwidth]{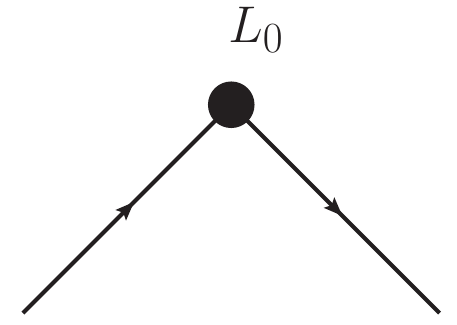}} \quad
   {\includegraphics[width=.15\textwidth]{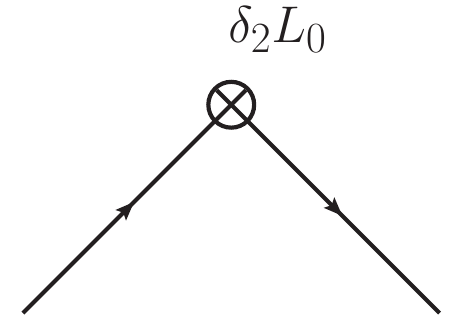}}\\
   \vspace{0.7 truecm}
   {\includegraphics[width=.15\textwidth]{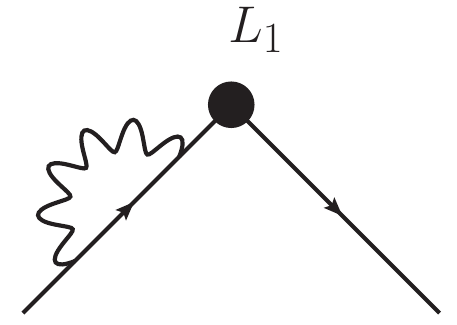}} \quad
   {\includegraphics[width=.15\textwidth]{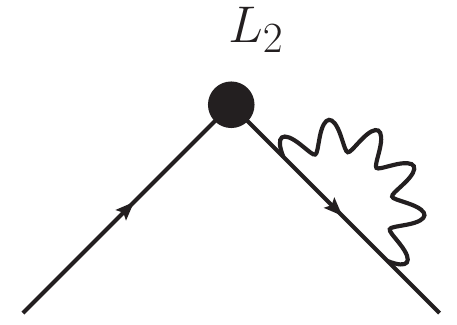}} \quad
   {\includegraphics[width=.15\textwidth]{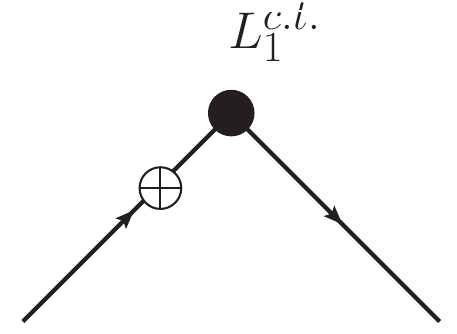}} \quad
   {\includegraphics[width=.15\textwidth]{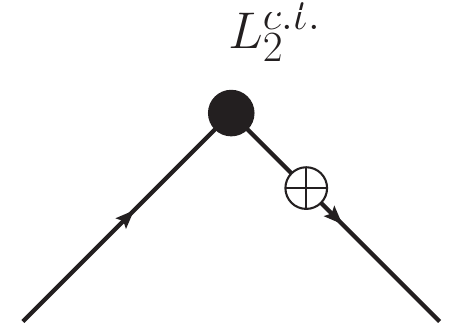}}  \quad\\
   \vspace{0.7 truecm}
   {\includegraphics[width=.15\textwidth]{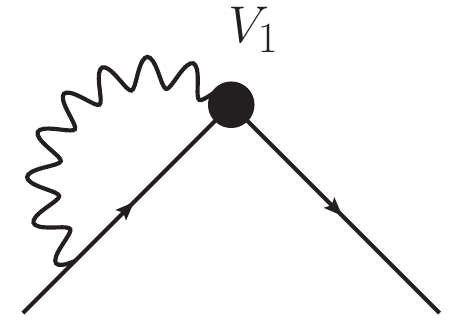}} \quad
   {\includegraphics[width=.15\textwidth]{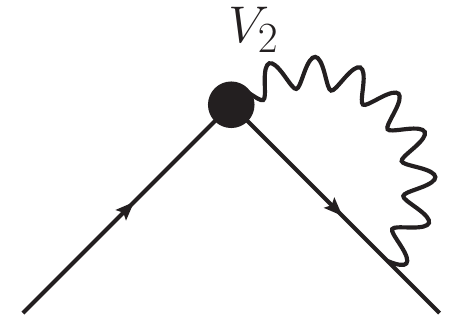}} \quad
   {\includegraphics[width=.15\textwidth]{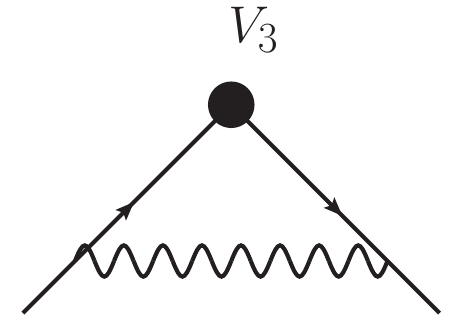}} \quad
   {\includegraphics[width=.15\textwidth]{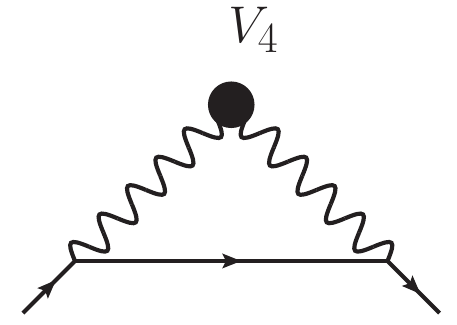}} 
   \caption{Feynman diagrams contributing to the electron EMT at $O(\alpha)$. The solid black dot represents the insertion of the EMT into the Green function, whereas a crossed dot indicates contributions from counter terms.} 
   \label{FeyDiag}
   \end{figure*}

The general parametrization of the electron and photon EMT matrix element between electron states can be written as~\cite{Kobzarev:1962wt, Pagels:1966zza, Ji:1996ek}
\begin{align}
&\langle e (p', s')| \, T^{\mu\nu}_i \, | e (p, s) \rangle \nonumber\\
& =
\bar u(p', s') \bigg( A_i (t) \, \frac{P^\mu P^\nu}{m_e} + J_i (t) \, \frac{iP^{\{\mu}\sigma^{\nu\}\rho}\Delta_\rho}{2m_e}
\nonumber \\
& \hspace{1.5cm} + D_i (t) \, \frac{\Delta^\mu \Delta^\nu - g^{\mu\nu} \Delta^2}{4m_e} + m_e \, \bar C_i (t) \, g^{\mu\nu} \bigg) u(p, s) \,,
\label{GeneralEMTParametrization_offF}
\end{align}
with $i=e,\gamma$ and $m_e$ the electron mass. 
In Eq.~\eqref{GeneralEMTParametrization_offF}, we use $P = \tfrac{1}{2}(p + p')$, $\Delta = p' - p$, $t = \Delta^2$, and $\sigma^{\mu\nu} = \tfrac{i}{2}(\gamma^\mu \gamma^\nu - \gamma^\nu \gamma^\mu)$.
The $A_i$, $J_i$, $D_i$ and $\bar C_i$ are the GFFs, which depend on $t$ and, in general, on the renormalization scale. 
(The latter dependence is suppressed to ease the notation.)
Below we will show results as a function of the dimensionless variable
\begin{equation}
\tau^2  = -\frac{t}{m_e^2} > 0 \,.
\end{equation}
Furthermore, the electron state in Eq.~\eqref{GeneralEMTParametrization_offF} obeys the covariant normalization
\begin{equation}
\braket{e(p', s') | e(p, s)} = 2p^0 \, (2\pi)^3 \, \delta_{s,s'} \, \delta(\vet{p}'-\vet{p}) \,,
\end{equation}
and the Dirac spinor satisfies $\bar u(p, s) \, u(p, s) = 2m_e$.

We are interested in the form factors $D_i(t)$ in Eq.~\eqref{GeneralEMTParametrization_offF} as obtained in perturbation theory. 
To this aim, we review the basic steps of the calculation that have been already presented in Ref.~\cite{Rodini:2020pis} in the context of considering the forward matrix elements of the EMT. 
We start by computing the Green function with the insertion of the EMT operator, that is,
\begin{align}
&\braket{e(p', s')|\, \mathrm{T}\qa T_i^{\mu\nu}(0)\exp\ta i\int d^4x \, \mathcal{L}_I(x) \tc\qc|e(p, s)}, \nonumber\\
&\text{with} \; 
\mathcal{L}_I = - e \, \bar \psi \slashed{A}\psi \,,
\label{emt_matrix_element}
\end{align}
where $\mathrm{T}$ indicates time ordering and the EMT is evaluated at the origin. 
(Choosing a different space-time point for the EMT would just lead to an irrelevant overall phase.)

The Feynman diagrams from the expansion of Eq.~\eqref{emt_matrix_element} up to $O(\alpha)$ are shown in Fig.~\ref{FeyDiag}. 
The diagram $L_0$ is the tree-level contribution, while $\delta_2 L_0$ is the overall vertex counterterm.
This counterterm coincides with the counterterm for the electron field because the total EMT is fully renormalized by means of Lagrangian renormalization and we are considering the matrix elements for the electron state only. 
The diagrams $L_{1,2}$ and $L_{1,2}^{c.t.}$ are the leg-loop corrections and the corresponding counterterm contributions, respectively.
We performed the calculations in the on-shell scheme, in which the sum $L_1+L_1^{c.t.}+L_2+L_2^{c.t.}$ vanishes. However, the total leg contribution $L_0(1+\delta_2)+L_1+L_1^{c.t.}+L_2+L_2^{c.t.}$ does not depend, at one-loop, on the renormalization scheme. This has been verified by performing the calculation also in the $\overline{\text{MS}}$ scheme.
Furthermore, the diagrams $V_{1,2}$ arise from the interaction term in $T_e^{\mu\nu}$, and $V_3$ is the one-loop electron vertex correction associated with the derivative term in $T_e^{\mu\nu}$.
Finally, $V_4$ represents the one-loop vertex correction where the photon is coupled to the external operator.
We note in passing that, at one loop, the extra term in the photon EMT in Eq.~(\ref{Tgamma}) proportional to $m_\gamma^2$ contributes to all the GFFs in Eq.~(\ref{GeneralEMTParametrization_offF}), except the form factors $D_i(t)$.

\section{Results for the form factor $D(t)$}
\label{FormFactorsEMTSection}
Since the EMT is conserved, the total form factor $D(t)$ is not renormalized, but the individual photon and electron contributions, generally, must be renormalized. 
However, at tree level the $D_i(t)$ vanish and, hence, they do not require any renormalization at one loop. 
On the other hand, the one-loop D-term has a divergence for vanishing momentum transfer in (standard) QED with massless photons~\cite{Berends:1975ah, Milton:1976jr, Donoghue:2001qc}.
More precisely, the origin of the divergence can be traced back to the photon contribution to the EMT.
Here we also show explicitly how a nonzero photon mass regulates this divergence.

The results for the form factors read
\begin{equation}
D_i(\tau^2 ,\lambda^2) = \int_0^1dx \int_0^{1-x}dy\frac{f_i(x,y)}{\tau^2 +a_i(x,y,\lambda^2)} \,,
\label{usefulDterm}
\end{equation}
where $\lambda =m_{\gamma}/m_e$, and 
\begin{align}
f_e(x,y) &= \frac{\alpha}{\pi}\frac{(x-2)(1-x-2y)^2}{y(1-x-y)} \,, \label{fe}\\
f_\gamma(x,y) &= \frac{\alpha}{\pi}\frac{1-x-(1+x)(1-x-2y)^2}{y(1-x-y)} \,, \\
a_e(x,y,\lambda^2) & = \frac{(1-x)^2+x\lambda^2}{y(1-x-y)} \,, \\
a_\gamma(x,y,\lambda^2) & = \frac{x^2+(1-x)\lambda^2}{y(1-x-y)} \,. \label{ag}
\end{align}
We have found complete numerical agreement between our result for the total form factor $D(\tau^2, \lambda^2 = 0)$ and the one reported in Refs.~\cite{Berends:1975ah, Milton:1976jr}.
The analytical result for the electron contribution in Eq.~\eqref{usefulDterm} is given by
\begin{align}
D_e(\tau^2 ,\lambda^2=0) = \frac{10 \, \alpha}{3 \pi \tau^2 } \bigg( 1 - \frac{\kappa}{2} \ln \frac{\kappa + 1}{\kappa -1} \bigg),
\, \,\kappa = \sqrt{1+\frac{4}{\tau^2 }} \,. \nonumber\\
\label{dTermE_exact}
\end{align}

In general, the value of the D-term is basically unconstrained.
It has been argued though that for any bound state the D-term should be negative --- see~\cite{Polyakov:2018zvc} and references therein.
One finds that for the electron in QED (with massless photons) the D-term is actually positive and infinite.
We repeat that this is due to the photon contribution to the EMT which, for $\tau^2 \to 0$, behaves as
\begin{equation}
D_\gamma(\tau^2 \ll 1, \lambda^2=0)\simeq \frac{\alpha\pi}{4\sqrt{\tau^2 }} \,. 
\label{dGdivD}
\end{equation}
In contrast, the electron contribution in Eq.~\eqref{dTermE_exact} at vanishing momentum transfer is finite,
\begin{equation}
D_e(\tau^2 =0, \lambda^2=0) = -\frac{5\alpha}{18\pi} \,.
\label{D-electron-dzero}
\end{equation}
For finite values of the photon mass we find
\begin{equation}
D_\gamma(\tau^2 =0, \lambda^2 \ll 1)\simeq \frac{\alpha}{3 \sqrt{\lambda^2}} \,,
\label{dGdivL}
\end{equation}
that is, a nonzero photon mass indeed leads to a finite D-term.
Obviously, the divergence $(+ \, \infty)$ is recovered in the limit $\lambda \to 0$.
The sum of the photon and electron contributions in Eqs.~\eqref{dGdivD} and~\eqref{D-electron-dzero} reproduces the limit for $\tau^2 \to 0$ of the total form factor $D(\tau^2)$ obtained in Refs.~\cite{Berends:1975ah, Milton:1976jr}.
We also notice that for finite values of $\tau^2 $ the form factors $D_i(\tau^2,\lambda^2 )$ are infrared safe.
\begin{figure}[t]
\centering
\includegraphics[width=0.45\textwidth]{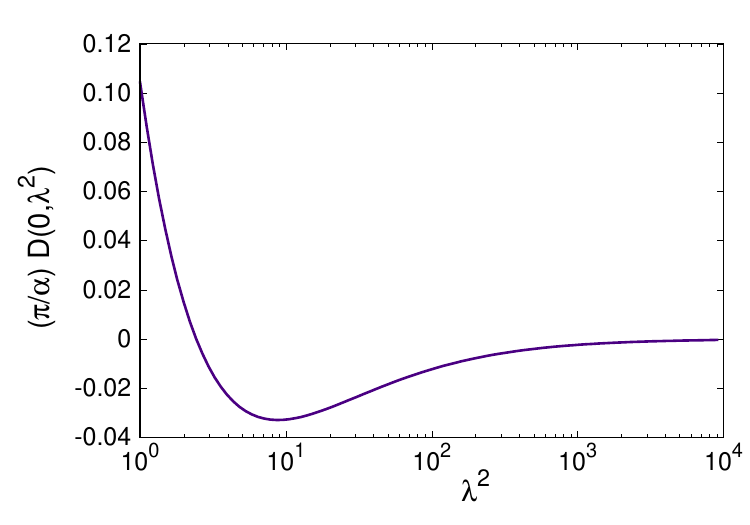}
\caption{The  total D-term of the electron as a function of $\lambda^2$.}
\label{dTermLambda}
\end{figure}
In Fig.~\ref{dTermLambda}, we show the D-term as a function of $\lambda^2$. 
It has a node at small values of $\lambda^2$, going from positive to negative values with increasing $\lambda^2$.   
In particular, there is a minimum at $\lambda^2_{\text{min}} \simeq 8.5$. 
For small values of the photon mass, the system behaves essentially as if the photon is massless: the D-term is positive and large as a result of the long-range photon-electron interaction. 
For very high values of the photon mass, the loop diagrams in Fig.~\ref{FeyDiag} are extremely suppressed, and the coupling of the (physical) electron to the EMT reduces to a contact interaction that behaves like the coupling at the tree-level where the D-term vanishes. 
But for moderate values of $\lambda^2$, the photon mass gives rise to a short-range interaction and the response of the physical electron to the coupling with the EMT mimics the behaviour of a bound state. 
However, this similarity does not imply that the electron-photon system becomes a bound state for a specific value/range of the photon mass. 

\begin{figure*}[t]
\centering
\subfloat[]{\includegraphics[width=0.48\textwidth]{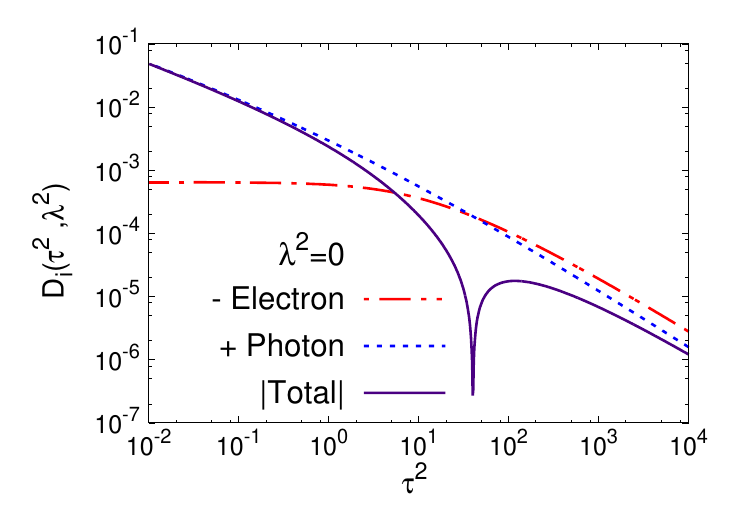}}\quad
\subfloat[]{\includegraphics[width=0.48\textwidth]{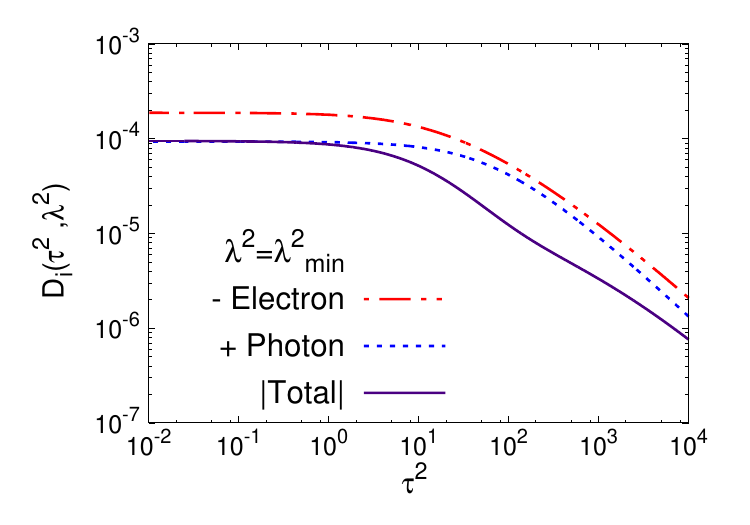}}
\caption{The form factor $D(\tau^2, \lambda^2)$ as a function of $\tau^2 $, for $\lambda^2=0$ (left) and for $\lambda^2=\lambda^2_{\text{min}}$ (right). 
The blue dotted curves are the photon contributions and the red dash-dotted curves correspond to the negative of the electron contributions. 
The solid purple curves represent the total form factor in absolute value.}
\label{dTerm_dm}
\end{figure*}

In Fig.~\ref{dTerm_dm}, we show the form factor $D(\tau^2, \lambda^2)$ as a functions of $\tau^2$, for a massless photon (panel (a)) and for $\lambda^2=\lambda^2_{\mathrm{min}}$ (panel (b)), by separating the terms due to the electron and the photon contributions to the EMT. 
For both values of $\lambda$, the photon contribution is always positive and the electron contribution always negative.
In the massless-photon case, the long-range electromagnetic interaction dominates at low $\tau^2$, whereas at intermediate $\tau^2 $ the negative electron contribution takes over so that the total form factor changes sign from positive to negative values.
The large-$\tau^2 $ behaviour of the total form factor is given by~\cite{Berends:1975ah}
\begin{equation}
D(\tau^2 \gg 1, \lambda^2=0) = \frac{\alpha}{\pi} \, \frac{4- \ln \tau^2}{\tau^2 } \,,
\label{high_d_BG}
\end{equation}
whereas from Eq.~\eqref{dTermE_exact} we can obtain the asymptotic behaviour of the electron contribution,
\begin{equation}
D_e(\tau^2 \gg 1, \lambda^2=0) = \frac{\alpha}{\pi} \, \frac{10-5 \ln \tau^2} {3\tau^2 } \,.
\label{high_d_el}
\end{equation}
By subtracting the electron contribution in Eq.~\eqref{high_d_el} from the total result in Eq.~\eqref{high_d_BG}, we find 
for the photon contribution
\begin{equation}
D_\gamma(\tau^2 \gg 1, \lambda^2=0) = \frac{\alpha}{\pi} \, \frac{2+2 \ln \tau^2}{3\tau^2 } \,.
\label{high_d_ph}
\end{equation}
This leads to the result
\begin{equation}
\lim_{\tau^2 \rightarrow \infty}\frac{D_e(\tau^2 , \lambda^2=0)}{D_\gamma(\tau^2 , \lambda^2=0)} = -\frac{5}{2} \,,
\label{ratio_asymptotic}
\end{equation}
which means that the photon and electron contributions vanish asymptotically with the same falloff, and the total form factor $D(\tau^2)$ approaches zero from negative values.
In the case of $\lambda^2=\lambda^2_{\text{min}}$, the negative electron contribution prevails over the positive photon contribution in the full range of the momentum transfer, and the two terms vanish asymptotically with the same falloff. 
The total form factor therefore resembles the features of a hadronic bound state, for which $D(t)$ is negative in the entire $t$-range~\cite{Polyakov:2018zvc}.

\section{Results for the pressure and shear-force distributions}
\label{PressureSection}
The distributions in the coordinate space of the pressure and shear forces can be obtained by Fourier transforming the form factors $D_i(\tau^2,\lambda^2)$~\cite{Polyakov:2002yz,Polyakov:2018zvc}.
Working in the Breit frame where
$P = (E,\bm{0}),$ and $ \Delta = (0,\bm{\Delta})$, we use $\bm{\tau}=\bm{\Delta}/m_e$ and its conjugate variable $\bm{\rho} = \bm{r} \, m_e$ to define the Fourier transform as
\begin{align}
\hat{D}_i(\rho,\lambda^2)  &= \int \frac{d^3\bm{\tau}}{(2\pi)^3}e^{-i\bm{\tau}\cdot\bm{\rho}}D_i(\tau^2, \lambda^2) \nonumber\\
&= \frac{1}{2\pi^2\rho} \, \text{FST} \ta \tau D_i(\tau^2,\lambda^2);\tau,\rho\tc, \label{dRhoGeneral}\\
\hat{\bar{C}}_i(\rho,\lambda^2) & = \int \frac{d^3\bm{\tau}}{(2\pi)^3}e^{-i\bm{\tau}\cdot\bm{\rho}}\bar{C}_i(\tau^2,\lambda^2) \,,
\end{align}
where  the Fourier sine transform is defined as
\begin{equation}
\text{FST}(f(x);x,y) \equiv \int_0^\infty dx \sin(xy)f(x) \,. 
\end{equation}
Using the expressions for the form factors in Eq.~\eqref{usefulDterm}, we find
\begin{equation}
\hat{D}_i(\rho,\lambda^2) = \int_0^1dx\int_0^{1-x}dy \, \frac{f_i(x,y)}{4\pi\rho} \, e^{-\rho\sqrt{a_i(x,y,\lambda^2)}} \,.
\label{dTermRho}
\end{equation}
We can then obtain the dimensionless pressure and shear distributions as~\cite{Polyakov:2018zvc}
\begin{align}
\hat{p}_i(\rho,\lambda^2) &= \frac{p_i(\rho,\lambda^2)}{m_e^4} = \frac{1}{6\rho^2}\frac{d}{d\rho}\rho^2\frac{d}{d\rho} \hat{D}_i(\rho,\lambda^2) - \hat{\bar{C}}_i(\rho,\lambda^2) \notag \\
& = \frac{1}{6}\frac{d^2}{d\rho^2}\hat{D}_i(\rho,\lambda^2)  + \frac{1}{3\rho}\frac{d}{d\rho}\hat{D}_i(\rho,\lambda^2) - \hat{\bar{C}}_i(\rho,\lambda^2) \,, 
\label{pressure_def}\\
\hat{s}_i(\rho,\lambda^2) &= \frac{s_i(\rho,\lambda^2)}{m_e^4} = -\frac{\rho}{4}\frac{d}{d\rho}\frac{1}{\rho}\frac{d}{d\rho} \hat{D}_i(\rho,\lambda^2) \nonumber\\
&= -\frac{1}{4}\frac{d^2}{d\rho^2}\hat{D}_i(\rho,\lambda^2)  + \frac{1}{4\rho}\frac{d}{d\rho}\hat{D}_i(\rho,\lambda^2) \,. 
\label{shear_def}
\end{align}

The contributions from the $\hat{\bar{C}}_i$ are proportional to $\delta(\rho) / \rho^2$, since it can be shown that the form factors $\bar{C}_i(\tau^2,\lambda^2)$ do not depend on $\tau^2$ (for any value of $\lambda^2$).
Specifically, we can write
\begin{equation}
\hat{\bar{C}}_i(\rho,\lambda^2) = \phi_i(\lambda^2) \, \frac{\delta'(\rho)}{\rho} \,, \; \text{with} \;\,
\phi_e(\lambda^2) = -\phi_\gamma(\lambda^2) \,,
\label{CbarHat}
\end{equation}
where we have used 
\begin{equation}
\int \frac{d^3\bm{\tau}}{(2\pi)^3}e^{-i\bm{\tau}\cdot\bm{\rho}} = - \frac{1}{2\pi \rho} \delta'(\rho) \,.
\end{equation}
Note that for the purpose of the present work one could use $\delta'(\rho) = - \delta(\rho)/ \rho$, with the understanding that $\int_0^\infty d\rho \, \delta(\rho) = \frac{1}{2}$.
In general, these two distributions are of course not identical.

Before presenting the numerical results in position space, we discuss some consistency checks that are mostly related to the conservation of the EMT. 
From the definitions in Eq.~\eqref{dRhoGeneral} and Eqs.~\eqref{pressure_def}--\eqref{shear_def}, we obtain
\begin{align}
\rho^2\hat{p}(\rho,\lambda^2) &=
\rho^2(\hat{p}_e(\rho,\lambda^2)+\hat{p}_\gamma(\rho,\lambda^2))\nonumber\\
& =
- \frac{\rho}{12\pi^2}\int_0^{\infty}d\tau \sin\ta \tau\rho\tc \tau^3D(\tau^2,\lambda^2) \,.
\end{align}
By integrating this relation over $\rho$ and interchanging the order of integration between $\rho$ and $\tau$, we find
\begin{equation}
\int_0^{\infty}d\rho\rho^2\hat{p}(\rho,\lambda^2) = \frac{1}{12\pi}\int_0^{\infty}d\tau \tau^3D(\tau^2,\lambda^2)\delta'(\tau)=0 \,,
\label{vonLaue}
\end{equation} 
where the last equality 
holds because, in the limit of $\tau\rightarrow 0 $,  $D_i(\tau^2,\lambda^2)$ goes like $1/\tau$ and therefore the derivative $d\ta\tau^3D_i(\tau^2,\lambda^2)\tc/d\tau$ is proportional to $\tau$  
--- see Ref.~\cite{Berends:1975ah}.
Equation~\eqref{vonLaue} represents one form of the von Laue condition, which follows from the conservation of the EMT~\cite{Polyakov:2018zvc,Lorce:2018egm}. 
If we neglect the form factors $\hat{\bar{C}}_i(\rho,\lambda^2)$, which are responsible for the fact that the individual ``parton'' EMTs are not conserved, then this condition holds for both the electron and photon contributions~\cite{Polyakov:2018zvc}, i.e.,
\begin{equation}
\hspace{-0.08 truecm}\int_0^{\infty}d\rho\rho^2\hat{p}_{i,D}(\rho,\lambda^2) = \frac{1}{12\pi}\int_0^{\infty}d\tau \tau^3D_i(\tau^2,\lambda^2)\delta'(\tau)=0 \,,
\label{vonLaue_partial}
\end{equation}
where $\hat{p}_{i,D}(\rho,\lambda^2)$ indicates the term 
in Eq.~\eqref{pressure_def} coming solely from the $\hat{D}_i(\rho, \lambda^2)$.
For the pressure and the shear distributions we can write
\begin{align}
\hat{p}_{i,\text{fin.}}(\rho,\lambda^2) &= \int_0^1dx\int_0^{1-x}dy \ e^{-\rho\sqrt{a_i(x,y,\lambda^2)}}\nonumber\\
&\times f_i(x,y)\frac{a_i(x,y,\lambda^2)}{24\pi\rho} \,, 
\label{pressure_res} \\
\hat{p}_{i,D,\text{sing.}}(\rho) &= 
\frac{\delta'(\rho)}{12\pi\rho}
\int_0^1dx\int_0^{1-x}dy \ f_i(x,y) \,, \label{pressure_res_sing}\\
\hat{s}_{i,\text{fin.}}(\rho,\lambda^2) &= -\int_0^1dx\int_0^{1-x}dy \ e^{-\rho\sqrt{a_i(x,y,\lambda^2)}}f_i(x,y)\nonumber\\
&\times \frac{ 3+3\sqrt{a_i(x,y,\lambda^2)}\rho+a_i(x,y,\lambda^2)\rho^2}{16\pi\rho^3} \,, 
\label{shear_res}\\
\hat{s}_{i,\text{sing.}}(\rho) &= 
\frac{3\delta(\rho)-\rho\delta'(\rho) }{8\pi\rho^2}
\int_0^1dx\int_0^{1-x}dy f_i(x,y) \,, \label{shear_res_sing}
\end{align}
where we have isolated the finite (fin.) and the singular (sing.) contributions. 
Therefore, not only the form factors $\bar{C}_i(\tau^2, \lambda^2)$ do give rise to singular terms at the origin but also the form factors $D_i(\tau^2, \lambda^2)$.
The von Laue condition in Eq.~\eqref{vonLaue_partial} is the result of an exact cancellation of the contributions of the finite and the singular terms.
A more general consequence of the conservation of the EMT is the relation~\cite{Polyakov:2018zvc}
\begin{equation}
\frac{2}{3}\frac{d \hat{s}_{i}(\rho,\lambda^2)}{d\rho} + \frac{2}{\rho}\hat{s}_{i}(\rho,\lambda^2) + \frac{d\hat{p}_{i,D}(\rho,\lambda^2)}{d\rho}=0 \,,
\label{general_cons}
\end{equation}
which is satisfied separately for the finite terms in Eqs.~\eqref{pressure_res},~\eqref{shear_res} and the singular terms in Eqs.~\eqref{pressure_res_sing},~\eqref{shear_res_sing}.
Finally, for $\lambda = 0$ the long-distance behaviour of the pressure and shear distributions is known~\cite{Donoghue:2001qc} --- see also the discussion in Ref.~\cite{Varma:2020crx}. 
Specifically, one has
\begin{align}
\hat{p}(\rho\rightarrow\infty,\lambda^2=0) &\simeq \hat{p}_\gamma(\rho\rightarrow \infty,\lambda^2=0) = \frac{\alpha}{24\pi\rho^4} + \ldots \,,  
\label{pressure_as} \\
\hat{s}(\rho\rightarrow \infty,\lambda^2=0) &\simeq \hat{s}_\gamma(\rho\rightarrow \infty,\lambda^2=0) = -\frac{\alpha}{4\pi\rho^4} + \ldots \,.
\label{shear_as}
\end{align}
We reproduce these results which are determined by the photon contribution to the EMT.
Numerically we find that 
the asymptotic limits are reached for $\rho \simeq 10^5$. 
\begin{figure*}[t]
\begin{center}
\subfloat[][]{\includegraphics[width=0.43\textwidth]{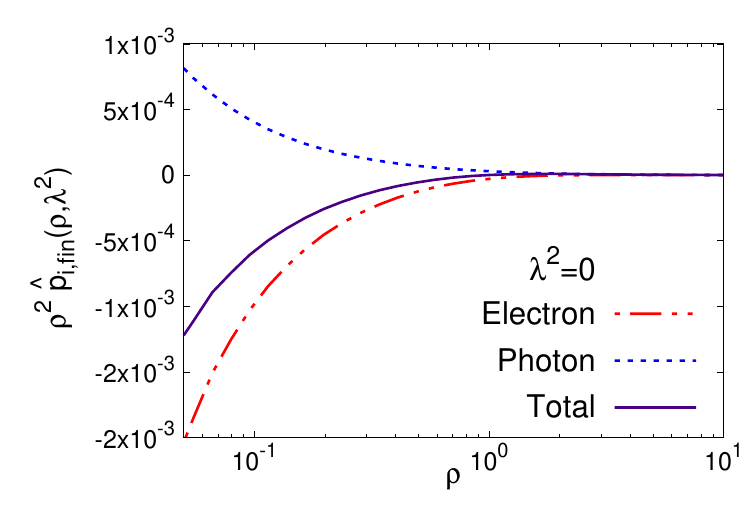}} \quad
\subfloat[][]{\includegraphics[width=0.43\textwidth]{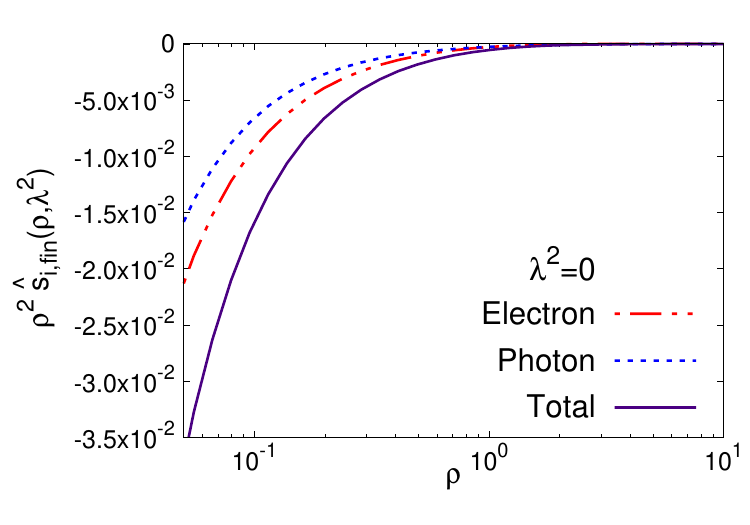}}\\
\subfloat[][]{\includegraphics[width=0.43\textwidth]{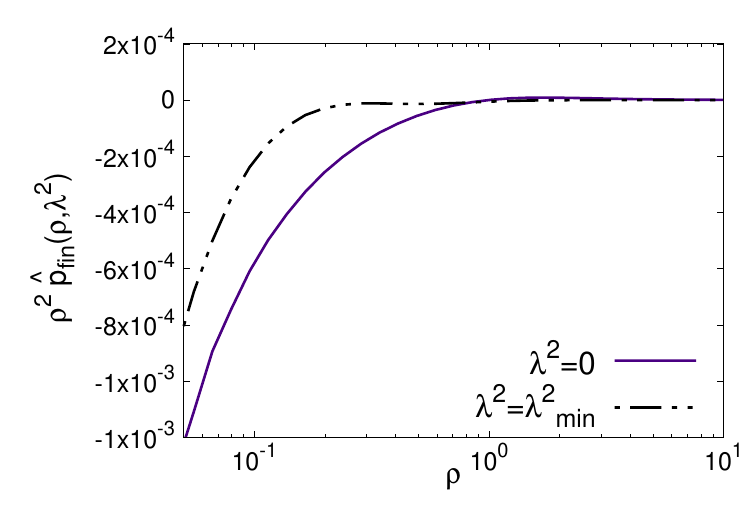}} \quad
\subfloat[][]{\includegraphics[width=0.43\textwidth]{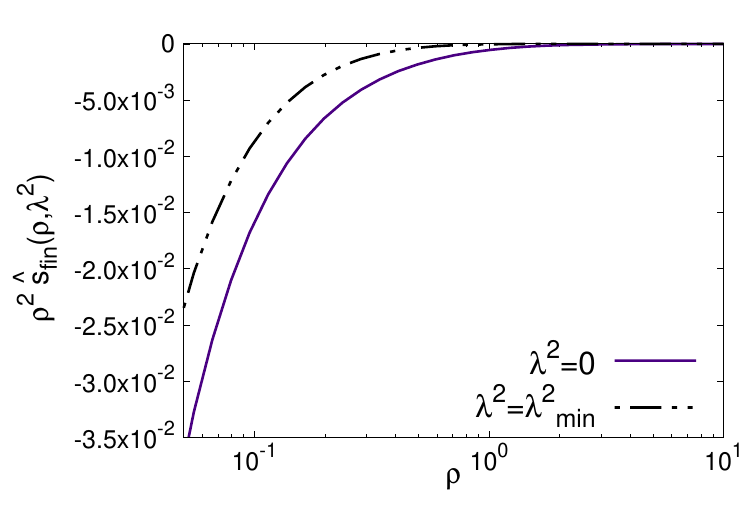}}
\end{center}
\caption{Finite contributions to the pressure and shear distributions as a function of $\rho$. 
Panels (a) and (b) show, respectively, the distributions in Eqs.~\eqref{pressure_res} and~\eqref{shear_res} multiplied by $\rho^2$, for $\lambda^2 = 0$.  
The blue dotted curves are the photon contributions and the red dash-dotted curves are the electron contributions, while the solid purple curves represent the total results.
Panels (c) and (d) show, respectively, the comparison of the pressure and shear distributions for $\lambda^2=0$ (solid purple curves) and $\lambda^2 = \lambda_{\mathrm{min}}^2$ (black dash-dotted curves).}
\label{fig_pressure_shear}
\end{figure*}

In Fig.~\ref{fig_pressure_shear}, we show the results for the finite contributions to the pressure and shear distributions as a function of $\rho$. 
Panels (a) and (b) are for a vanishing photon mass ($\lambda^2 = 0$), while panels (c) and (d) contain the comparison between $\lambda^2 = 0$ and $\lambda^2 = \lambda_{\text{min}}^2$.
The total (finite) pressure distributions are always negative, due to the negative contributions from the electron part of the EMT.
In the case of hadrons, the pressure distributions are (also) negative (attractive) at large distances but positive (repulsive) at small distances --- see~\cite{Polyakov:2018zvc} and references therein. 
For the electron in QED a positive contribution to the pressure comes only in the form of a singular term at the origin.
This difference between a hadronic bound state and the electron is caused by the different behaviour of the GFFs at large momentum transfer.
In our calculation, the finite contribution to the distribution of the shear force is negative, regardless of the value of the photon mass, while a singular contribution is located at the origin.
In contrast, for hadrons the shear distributions tend to be positive for any finite distance.
Because of the finite interaction range, both the pressure and shear distributions for finite photon masses vanish faster at large distances than for $\lambda = 0$.

\begin{figure*}[t]
\begin{center}
\subfloat[][]{\includegraphics[width=0.48\textwidth]{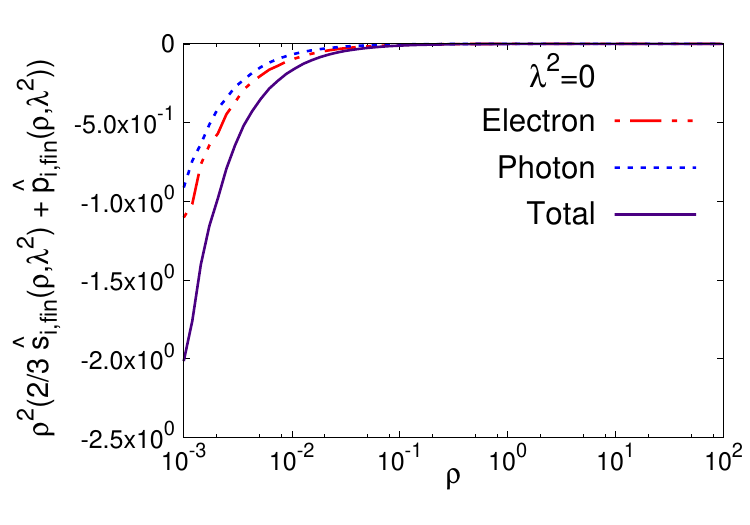}}\quad
\subfloat[][]{\includegraphics[width=0.48\textwidth]{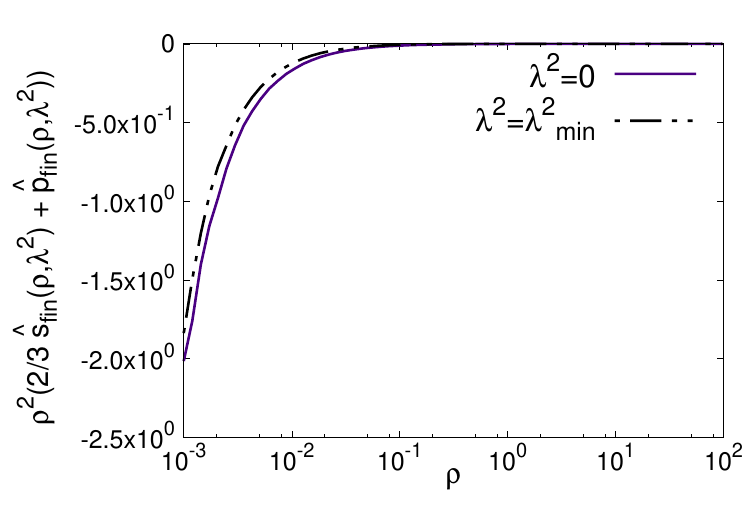}}\\
\subfloat[][]{\includegraphics[width=0.48\textwidth]{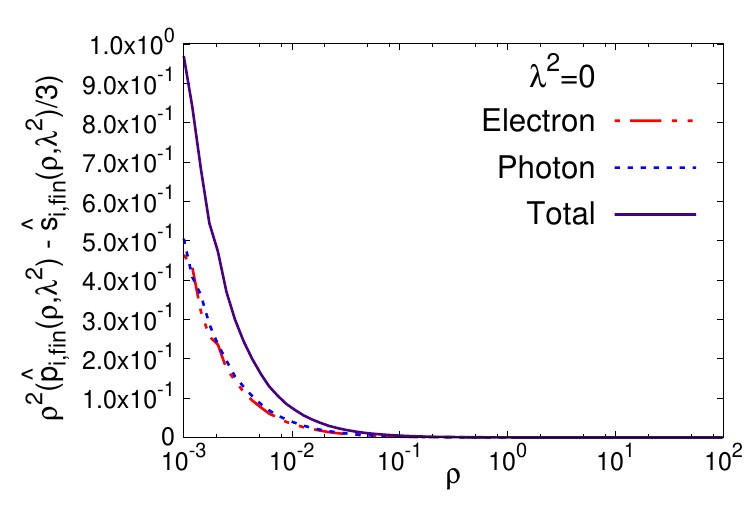}}\quad
\subfloat[][]{\includegraphics[width=0.48\textwidth]{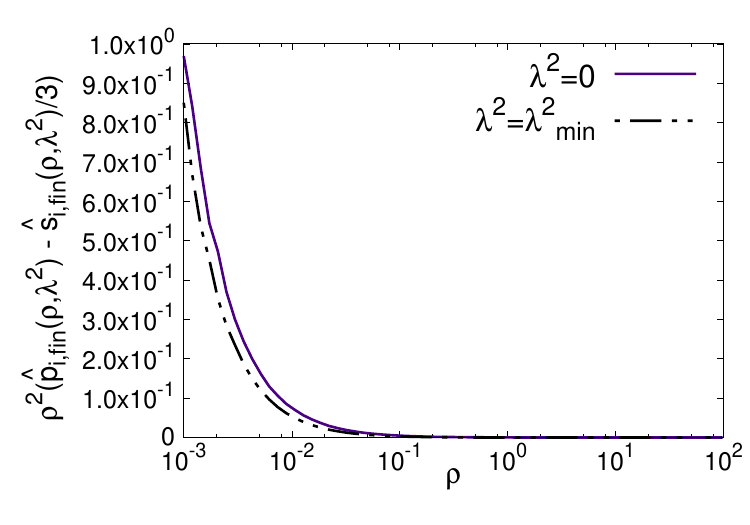}}\\
\end{center}
\caption{Finite contributions to the normal and tangentail forces in Eqs.~\eqref{normal_force},~\eqref{tangential_force} as a function of $\rho$. 
The blue dotted curves are the photon contributions and the red dash-dotted curves are the electron contributions, while the solid purple curves represent the total results.
Panels (b) and (d) compare the total forces for $\lambda^2=0$ (solid purple curves) and for $\lambda^2 = \lambda_{\mathrm{min}}^2$ (black dash-dotted curves).}
\label{fig_forces_norm}
\end{figure*}

The pressure and shear distributions allow one to define normal and tangential forces inside a system~\cite{Polyakov:2018zvc}.
Specifically, the forces experienced by a spherical shell of radius $\rho$ are given by
\begin{align}
F_n &= 4 \pi \rho^2 \bigg( \frac{2}{3} \hat{s}(\rho, \lambda^2) + \hat{p}(\rho, \lambda^2) \bigg) \,,
\label{normal_force} \\
F_t &= 4 \pi \rho^2 \bigg( \hat{p}(\rho, \lambda^2) - \frac{1}{3} \hat{s}(\rho, \lambda^2) \bigg) \,, 
\label{tangential_force}
\end{align}
and the corresponding numerical results (without the factor $4\pi$) are shown in Fig.~\ref{fig_forces_norm}.
We find that for nonzero $\rho$ the normal force $F_n$ is negative and the tangential force $F_t$ is positive, while both forces have a singular contribution at the origin.
For comparison, the normal force in hadrons is positive, and the tangential force distribution typically switches sign.
\begin{figure*}
\begin{center}
\subfloat[][]{\includegraphics[width=0.48\textwidth]{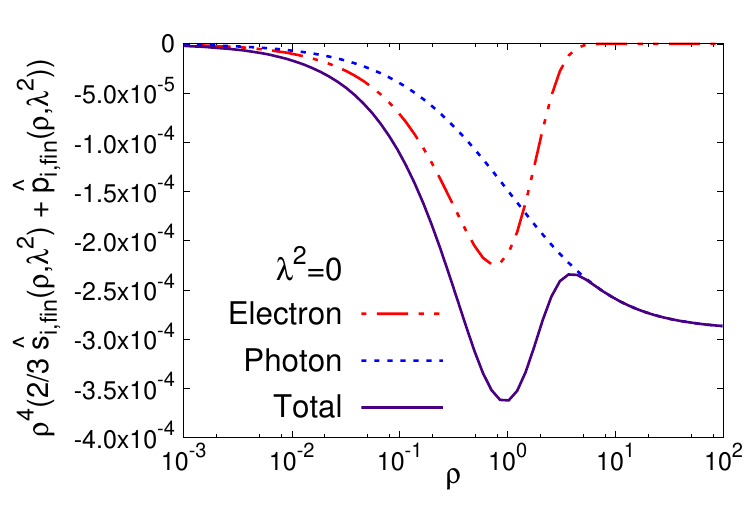}} \quad
\subfloat[][]{\includegraphics[width=0.48\textwidth]{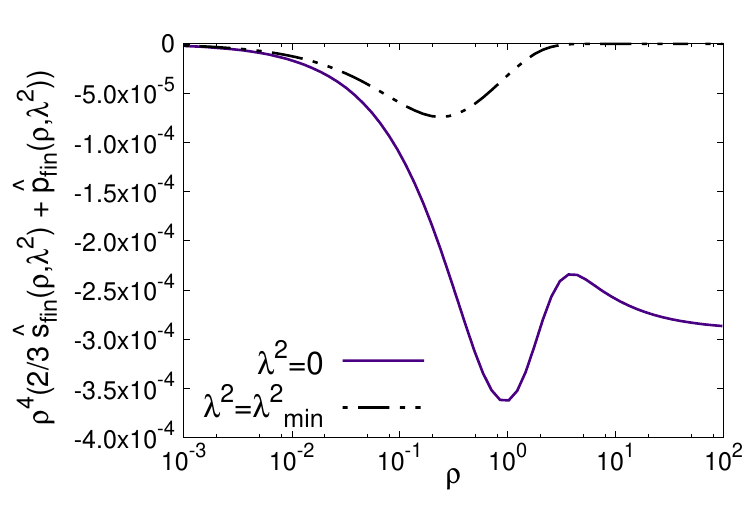}}
\end{center}
\caption{Integrand for numerator of the mechanical radius in Eq.~\eqref{mech_radius} as a function of $\rho$, for both $\lambda^2 = 0$ (panel (a)) and $\lambda^2 = \lambda_{\text{min}}^2$ (panel (b)).
(Note that the singular terms in the pressure and shear distributions do not contribute to this integral.)
The blue dotted curve shows the photon contribution, the red dash-dotted curve the electron contribution, the solid purple curves the total result for $\lambda^2 = 0$, and the black dash-dotted curve the total result for $\lambda^2 = \lambda_{\text{min}}^2$.
}
\label{fig_numerator_radius}
\end{figure*}

The normal force in Eq.~\eqref{normal_force} was used to define the so-called mechanical radius of a system according to~\cite{Polyakov:2018zvc}
\begin{equation}
\langle r^2 \rangle_{\text{mech}} \equiv \frac{\int d^3r \, r^2 \, \big[\frac{2}{3} \hat{s}(r) + \hat{p}(r) \big]}{\int d^3r \, \big[ \frac{2}{3} \hat{s}(r) + \hat{p}(r) \big]} \,.
\label{mech_radius}
\end{equation}
This can be re-expressed through the form factor $D(t)$~\cite{Polyakov:2018zvc}, where for the electron (with the dimensionless variable $\tau^2$) one has
\begin{equation}
\langle r^2(\lambda^2) \rangle_{\text{mech}} = \frac{6 \, D(0, \lambda^2)}{m_e^2 \int_0^\infty d \tau^2 D (\tau^2, \lambda^2)} \,.
\label{mech_radius_lambda}
\end{equation}
For $\lambda^2 = 0$, the numerator as well as the denominator in Eq.~\eqref{mech_radius_lambda} are undefined:
as discussed above, the D-term is infinite, while the integral in the denominator diverges at both the lower and the upper integration limits.
The same conclusion follows from the definition of the mechanical radius in Eq.~\eqref{mech_radius} based on distributions in position space.
In that case the numerator diverges due to the $1/\rho^4$-behaviour of the integrand at large distances (see Eqs.~\eqref{pressure_as} and~\eqref{shear_as}), whereas the denominator diverges due to a singularity for $\rho \to 0$ in the integral of the shear distribution.
On the other hand, for any $\lambda^2 \neq 0$, Eq.~\eqref{mech_radius_lambda} provides $\langle r^2(\lambda^2 \neq 0) \rangle_{\text{mech}} = 0$, since the D-term is finite but the denominator (still) diverges because of a singularity for $\rho \to 0$.
(In this context see also Fig.~\ref{fig_numerator_radius} which shows the integrand for the numerator in Eq.~\eqref{mech_radius} for both $\lambda^2 = 0$ and $\lambda^2 = \lambda_{\text{min}}^2$.
In the latter case, the pressure and shear distributions fall off much faster at large $\rho$, leading to a finite value for the numerator in Eq.~\eqref{mech_radius}.)
These results should also hold for arbitrary order in perturbation theory.
It is therefore tempting to define the mechanical radius of the electron in QED (with a massless photon) according to 
$\langle r^2 \rangle_{\text{mech}}^{e^-} \equiv \lim_{\lambda \to 0} \langle r^2(\lambda^2 \neq 0) \rangle_{\text{mech}} = 0$.
However, we do not assign much significance to this result.
We rather conclude that the concept of the mechanical radius cannot be applied for systems whose form factor $D(t)$ does not drop fast enough at large momentum transfer --- see also the corresponding discussion in Ref.~\cite{Polyakov:2018zvc}.

\section{Conclusions}
\label{SummarySection}
The off-forward matrix elements of the EMT, which are parametrized through several GFFs, encode a wealth of information about the energy, spin, pressure and shear distributions inside a particle. 
We focused our attention on the one-loop QED calculation of the GFF $D(t)$ for an electron, by separately evaluating the contributions from the electron and the photon parts of the EMT.
The form factor $D(t)$, which represents a fundamental quantity of the (physical) electron, contains the information about the (total) pressure and shear distributions.
The D-term, that is $D(t = 0)$, is infinite, where the infinity is caused by the photon contribution to the EMT in combination with the long-range Coulomb interaction --- see also Ref.~\cite{Donoghue:2001qc}.
On the other hand, for a nonzero (and sufficiently large) photon mass one finds that the D-term becomes negative and, in fact, the form factor $D(t)$ is negative for the entire $t$-range.
The same qualitative results hold for hadronic bound states~\cite{Polyakov:2018zvc}.
Therefore the ``physical'' electron, which is composed of a bare electron and a massive photon, mimics a bound state.
However, there is one crucial difference:
For both a zero and nonzero photon mass the form factor $D(t)$ essentially drops like $1/|t|$ at large momentum transfer.
This is (considerably) slower than for a hadronic bound state~\cite{Polyakov:2018zvc, Lorce:2018egm}. 
When making this comparison between the electron and hadrons, we do not consider electromagnetic contributions to $D(t)$ for a hadron, which do actually also lead to a $1/|t|$ behavior at large $|t|$ as discussed recently in Ref.~\cite{Varma:2020crx}.
As a result, the pressure and shear distributions in position space show significant qualitative differences in the two cases.
In particular, for the electron in QED both distributions exhibit a delta function singularity at the origin $r = 0$, which is not known for hadrons.
We repeat that the GFFs, including $D(t)$, are fundamental quantities and, in general, differences between the electron and hadronic bound states are due to the long-range QED effects and the behavior of the GFFs at large momentum transfer.


\section*{Acknowledgments}

The work of A.M.~was supported by the National Science Foundation under the Grant No.~PHY-1812359, and by the U.S. Department of Energy, Office of Science, Office of Nuclear Physics, within the framework of the TMD Topical Collaboration.
The work of B.P.~and S.R.~is part of a project that has received funding from the European Union's Horizon 2020 research and innovation programme under grant agreement STRONG - 2020 - No~824093.



\end{document}